\documentclass[fleqn,usenatbib]{mnras}
\usepackage{newtxtext,newtxmath}
\usepackage{booktabs}
\usepackage{siunitx}
\usepackage{nicefrac}
\DeclareSIUnit\parsec{pc}
\usepackage[T1]{fontenc}
\DeclareRobustCommand{\VAN}[3]{#2}
\let\VANthebibliography\thebibliography
\def\thebibliography{\DeclareRobustCommand{\VAN}[3]{##3}\VANthebibliography}

\usepackage{graphicx}	
\usepackage{amsmath}	

\title[Jeans analysis of Gaia DR3 K-dwarfs]{Local dark matter density from Gaia DR3 K-dwarfs using Gaussian processes}

\author[Söding et al.]{
Laurin Söding$^{1}$\thanks{E-mail: soeding@physik.rwth-aachen.de},
Ruben L. Bartel$^{1}$,
and Philipp Mertsch$^{1}$
\\
$^{1}$Institute for Theoretical Particle Physics and Cosmology, RWTH Aachen University, Sommerfeldstr. 16, 52074 Aachen, Germany
}

\date{Accepted XXX. Received YYY; in original form ZZZ}
\pubyear{\the\year{}}

\newcommand*\diff{\mathop{}\!\mathrm{d}} 
\begin{document}
\label{firstpage}
\pagerange{\pageref{firstpage}--\pageref{lastpage}}
\maketitle

\begin{abstract}
    The local dark matter density provides constraints on dark matter models and is of importance for experiments hoping to detect dark matter particles in the laboratory.
    The advent of extensive survey data calls for more complex physical modelling and more sophisticated statistical analysis, particularly to account for correlated uncertainties.
    In this paper, we perform a vertical Jeans analysis, including a local approximation of the tilt term, using a sample of $200\,000$ K-dwarf stars from the Gaia DR3 catalogue.
    After combination with the Survey-of-Surveys (SoS) catalogue, $160\,888$ of those have radial velocity measurements.
    We use Gaussian processes as priors for the covariance matrix of radial and vertical velocities.
    Joint inference of the posterior distribution of the local dark matter density and the velocity moments is performed using geometric variational inference.
    We find a local dark matter density of ${\rho_\mathrm{dm} = 0.0117 \pm 0.0035\, \mathrm{M}_\odot\,\mathrm{pc}^{-3} = 0.44 \pm 0.13\, \mathrm{GeV}\,\mathrm{cm}^{-3}}$ at the Sun's position, which is in agreement with most other recent analyses.
    By comparing a ($z$-dependent) Gaussian process prior with a ($z$-independent) scalar prior for the tilt term, we quantify its impact on estimates of the local dark matter density and argue that careful modelling is required to mitigate systematic biases.
\end{abstract}

\section{Introduction}
Most of the mass in the Milky Way is believed to be in the form of dark matter (DM).
While a non-gravitational detection is still lacking, the gravitational effects on the dynamics and evolution of Galaxies are sizeable and can be quantified with ever-increasing accuracy.

In recent years, the Gaia mission \citep{GaiaDR3} has presented unprecedented data on the positions and velocities of stars in the nearby region of the Milky Way.
This wealth of information has opened up new avenues for studying the structure of the Milky Way and the distribution of dark matter within it, as assumptions that had to be made in the past can now be informed by data.

The increase in data quantity has been accompanied by an increase in the complexity of the models used to analyse and interpret it.
Simple fits of extremely symmetric, parametric models are no longer sufficient to capture the complexity of the data \citep[][]{Wang2017,Antoja2018,Sivertsson2022,Rehemtulla2022,Li2025}.
To properly estimate statistical uncertainties, Bayesian inference has emerged as the method of choice.
However, tracing the posterior distribution, for example using Markov Chain Monte Carlo (MCMC) methods, is quickly becoming prohibitively expensive as the number of parameters increases.

Another way of performing Bayesian inference is variational inference, that is, approximating the posterior distribution with a simpler, parametric distribution.
In recent years, new methods based on information field theory have been developed to perform variational inference in high-dimensional parameter spaces \citep{MGVI,geoVI}.
With these methods, the prior space is not restricted to a few scalar parameters, but can easily be extended to thousands of latent parameters describing e.g. physical fields on a fine grid.

Extracting the local dark matter density from the data is a non-trivial task and there are many methodological approaches for this.
First attempts to model the local matter density using stellar kinematics date back at least a century \citep[e.g.][]{Kapteyn1922,Oort1932,Bahcall1984}.
As the main observable quantity at our disposal are stellar positions and velocities, most methods focus on the kinematics of some stellar population.
For example, there are recent methods based on an analysis of circular velocities \citep[e.g.][]{Salucci2010,Nesti2013,Pato2015_RotCurve,Huang2016_RotCurve,Benito2019_RotCurve,Karukes2019_RotCurve,Lin2019_RotCurve,deSalas2019_RotCurve,Ablimit2020_RotCurve,Benito2021_RotCurve,Sofue2020_RotCurve,Zhou2023_RotCurve,Ou2024_RotCurve,McMillan2017_RotCurve,Cautun2020_RotCurve}, fitting a parametric distribution function \citep[e.g.][]{Bienayme2014_DFF,Piffl2014_DFF,Binney2015_DFF,Cole2017_DFF}, normalising flows \citep{Lim2025_NF}, halo star mass models \citep[e.g.][]{Kafle2014_HS,Hattori2021_HS,Wegg2019_HS}, Jeans anisotropic modelling \citep[e.g.][]{Nitschai2020_aniJeans,Nitschai2021_aniJeans}, modelling of phase-space-spirals \citep[e.g.][]{Widmark2021_PSS,Guo2022_PSS}, very local analyses \citep[e.g.][]{Holmberg2000_Local,Schutz2018_Local,Buch2019_Local}, and vertical Jeans analyses \citep[e.g.][]{Garbari2012,McKee2015_vJeans,Xia2016_vJeans,Hagen2018_vJeans,Sivertsson2018_vJeans,Guo2020_vJeans,Salomon2020_vJeans,Wardana2020_vJeans,Cheng2024}.
For a more comprehensive overview and list of previous works, we refer to a review by \citet{deSalas2021}.
While most analyses agree on a local dark matter density on the order of $0.01\,\mathrm{M_\odot\,pc^{-3}}$ at the Sun's position, they differ in their methodological approach and the level of rigour in their treatment of uncertainties.

In this analysis, we use a vertical Jeans analysis to estimate the local dark matter density.
We use stellar data from mainly the Gaia DR3 catalogue \citep{GaiaDR3} and extend it with data from the Survey-of-Surveys (SoS) catalogue \citep{Tsantaki2022}.
Doing so, we focus on main-sequence stars which are abundant enough to yield good statistics in velocity and density space.
Setting us apart from previous analyses, we use a non-parametric model for the velocity moments in terms of Gaussian processes and describe correlations between vertical and radial motion of stars as a function of height above and below the disk.
This allows us to model the velocity moments in a flexible way without having to assume a specific functional form, while naturally propagating various uncertainties.
As a result of our Bayesian inference, we obtain samples tracing the posterior probability distribution not only of the local dark matter density but also of the velocity moments.
We quantify the effect of the so-called tilt term in the Jeans equation and find that it is subdominant for our analysis.

The Jeans equation connects the density profile of a stellar population and their velocity moments to the gravitational potential of the Milky Way \citep[][]{Binney2008}.
The gravitational potential is then connected to the total mass density of the Milky Way by means of the Poisson equation.
The total mass density can then be separated into a visible and a dark matter component.

The paper is structured as follows: In section \ref{sec:Data}, we present the data used in this analysis, including the disc mass model and the selection of our tracer population.
In section \ref{sec:Methods}, we present the methods used including the Poisson-Jeans system of equations, the variational inference method, and our forward model connecting the data to the parameters.
In section \ref{sec:Results}, we present the results of our analysis, including the local dark matter density and the velocity moments, before we conclude in section \ref{sec:Conclusions}.

\section{Data}
\label{sec:Data}
In recent times, thanks to the Gaia mission, datasets containing 6D stellar measurements, that is 3D positions and 3D velocities, have grown significantly.
For a Jeans analysis, such data is crucial as it constrains the velocity statistics which are related to the gravitational potential of the Milky Way.
If one knew the gravitational potential, constraining the dark matter component would be a problem of component separation.
We will make use of the Gaia DR3 catalogue \citep{GaiaDR3} extensively.
We extend the Gaia catalogue with data from the Survey-of-Surveys (SoS) catalogue \citep{Tsantaki2022} to obtain a larger sample of stars with 3D velocity information.
For the reference, we will fix some general numbers that are used for coordinate transformations, noting that their exact values are not important for the results of this analysis.
We assume a height of the Sun above the Galactic disk of $z_\odot = 14.5\,\mathrm{pc}$ \citep{Skowron2019}, a Solar radius of $R_\odot = 8.15\,\mathrm{kpc}$ \citep{Reid2019}, a circular velocity of the local standard of rest of $\varv_\odot = 232\,\mathrm{km\,s^{-1}}$ \citep{McMillan2017_RotCurve}, and a peculiar motion of the Sun of $(U_\odot, V_\odot, W_\odot) = (11.1, 12.24, 7.25)\,\mathrm{km\,s^{-1}}$ \citep{Schoenrich2010}.

In the following, we will shortly present the disk mass model used in section \ref{sec:DiscMassModel}.
Then, we will discuss the data selection process for our tracer population and how we extract the density profile in section \ref{sec:StellarTracerPopulation}.
We finish this section by discussing our data fusion with the SoS catalogue in section \ref{sec:StellarVelocities}.

\subsection{Disc mass model}
\label{sec:DiscMassModel}
For separating the dark matter density from other gravitating masses, we need to model those other components.
We use the same model as \citet{Garbari2012}, that is we use estimates of the mid-plane densities from \citep{Flynn2006} with uncertainties as proposed by \citep{Garbari2012}.
For convenience, we list those values in Table~\ref{tab:massmodel}.
For the $z$-dependence of the density components, we assume an isothermal profile, that is
\begin{equation}
    \label{eq:IsothermalMassModel}
    \rho_s(z) = \sum_j \rho_{s,0}^j \exp\left(-\frac{\Phi(z)}{\overline{(\varv_{s,0}^j)^2}}\right)\,,
\end{equation}
where $\rho_s$ is the total density of visible mass, $\rho_{s,0}^j$ is the mid-plane density of component $j$, $\Phi$ is the gravitational Potential of the Milky Way, and $\overline{(\varv_{s,0}^j)^2}$ is the effective velocity dispersion of component $j$.

\begin{table}
    \centering
    \begin{tabular}{l|cc}
        \toprule
        Component         & $\rho_{s,0}^j$ $[\mathrm{M}_{\odot}\,\mathrm{pc}^{-3}]$ & $\overline{(\varv_{s,0}^j)^2}^{1/2}$ $[\mathrm{km}\,\mathrm{s}^{-1}]$ \\
        \midrule
        H$_2$*            & 0.021                                                   & 4.0 $\pm$ 1.0                                                         \\
        HI(1)*            & 0.016                                                   & 7.0 $\pm$ 1.0                                                         \\
        HI(2)*            & 0.012                                                   & 9.0 $\pm$ 1.0                                                         \\
        Warm gas*         & 0.0009                                                  & 40.0 $\pm$ 1.0                                                        \\
        Giants            & 0.0006                                                  & 20.0 $\pm$ 2.0                                                        \\
        $M_V < 2.5$       & 0.0031                                                  & 7.5 $\pm$ 2.0                                                         \\
        $2.5 < M_V < 3.0$ & 0.0015                                                  & 10.5 $\pm$ 2.0                                                        \\
        $3.0 < M_V < 4.0$ & 0.0020                                                  & 14.0 $\pm$ 2.0                                                        \\
        $4.0 < M_V < 5.0$ & 0.0022                                                  & 18.0 $\pm$ 2.0                                                        \\
        $5.0 < M_V < 8.0$ & 0.007                                                   & 18.5 $\pm$ 2.0                                                        \\
        $M_V > 8.0$       & 0.0135                                                  & 18.5 $\pm$ 2.0                                                        \\
        White dwarfs      & 0.006                                                   & 20.0 $\pm$ 5.0                                                        \\
        Brown dwarfs      & 0.002                                                   & 20.0 $\pm$ 5.0                                                        \\
        Thick disc        & 0.0035                                                  & 37.0 $\pm$ 5.0                                                        \\
        Stellar halo      & 0.0001                                                  & 100.0 $\pm$ 10.0                                                      \\
        \bottomrule
    \end{tabular}
    \caption{Parameters of the visible components $j$ of the disc mass model: local volume mass density in the galactic mid-plane $\rho_{s,0}^j$ and vertical velocity dispersion in the galactic mid-plane $\overline{(\varv_{s,0}^j)^2}$. All gaseous components (indicated with *) are assumed to have uncertainties of $50\%$ on their densities and the stellar components have $20\%$. Model taken from \citet{Flynn2006}}
    \label{tab:massmodel}
\end{table}

Several alternative Galactic models have been developed in the literature, including for example the Besançon model \citep{Robin2003_Besancon,Robin2022_Besancon} and the Galactic model of \citet{McMillan2017_RotCurve}.
However, the accuracy of any such model remains challenging to evaluate for the following two reasons:
First, these models assume a parametric, often highly symmetric Galaxy, whereas -- as will become apparent below -- the Milky Way exhibits significant asymmetries and inhomogeneities.
Second, the observational data available to constrain model parameters are incomplete, as certain stellar populations are difficult to observe directly and gaseous components remain undetected or poorly characterised (e.g. the value of the conversion factor from CO to H$_2$).
We do not expect to gain much by using a more complex model in this analysis.
While there are undoubtedly improvements in modern models, they do not sufficiently address the most pressing issue of describing a non-ideal system.
Therefore, we will use the model from \citet{Flynn2006} as it allows for a direct comparison of our results with previous works.
We will however incorporate some newer analysis of the stellar and gaseous components in the form of surface mass densities within $1.1\,\mathrm{kpc}$ of the Galactic mid-plane by \citet{McKee2015_vJeans}.
They estimate ${\Sigma_{\mathrm{g},1.1}=12.6\pm1.6\,\mathrm{M}_\odot\,\mathrm{pc}^{-2}}$ for the gaseous component and ${\Sigma_{*,1.1}=31.2\pm3.0\,\mathrm{M}_\odot\,\mathrm{pc}^{-2}}$ for the stellar component.

\subsection{Stellar tracer population}
\label{sec:StellarTracerPopulation}
The stellar population to be analysed necessarily has to fulfil the following two criteria: First, it has to be long-lived so that it is in dynamical equilibrium with the gravitational potential of the disc.
Second, it has to be bright enough so that volume-complete measurements are available up to sufficiently large distances above and below the Sun's position in the Galactic disc.
To this end, we make use of the Gaia DR3 catalogue \citep{GaiaDR3}, selecting stars within a cylinder of radius $200\,\mathrm{pc}$ and height $|z|\leq 1800\,\mathrm{pc}$ as computed using the mean estimated parallax, corrected according to \citet{Lindgren2021}. 
Using the mean parallax here omits some stars whose true parallax would lie inside the region of interest, whereas its estimated mean does not.
As we aim to achieve sufficient volume completeness, we will address this later by performing additional cuts.
The instruments of Gaia are limited in apparent magnitude both towards very bright and very faint stars.
In \citet{Fabricius2021}, the completeness of Gaia data was investigated by comparison to other surveys and models.
For our purposes, completeness can be assumed approximately for apparent magnitudes in G-band between $G\geq 12\,\mathrm{mag}$ and $G\leq 17\,\mathrm{mag}$.
Especially the faint limit on completeness is ill-defined and strongly depends on position and crowding.
Since magnitudes as faint as $G\approx 21\,\mathrm{mag}$ can be detected, the upper limit above should be understood as a conservative estimate.
An apparent magnitude lower (upper) limit, together with a minimum (maximum) distance we want to consider yields the lower (upper) limit on the absolute magnitude.
Choosing these limits too tight (lenient) leads to a loss of statistics due to restrictions in distance (magnitude) space.

As a compromise, we restrict our analysis to stars with an absolute magnitude $M_G\in[6.0, 7.4]$, which, for main sequence stars, falls into the K-dwarf category.
Using the apparent magnitude limits from above, this corresponds approximately to a distance range of $z\in[200, 1000]\,\mathrm{pc}$.
We look at another quantity as an indicator of completeness, that is the quality of parallax measurements.
The fraction of stars with a parallax error exceeding $20\,\%$ increases sharply from almost $0\%$ up to $|z|\approx 1200\,\mathrm{pc}$ to $70\%$ at $|z|\approx 1800\,\mathrm{pc}$, which could indicate a loss of data quality in this regime.
We choose this value as our upper limit and note that the following results in this paper are insensitive to small changes in this value.
This leaves us with $464\,172$ stars, of which only $4\,123$ have no colour (\texttt{g\_rp}) information (mostly at small distances where they are negligible in number).

Next, we clean our sample by removing objects that do not lie on the main sequence.
Above the main sequence, we expect to find e.g. unresolved binaries, which would typically appear brighter; and stars with an unusually high metallicity.
Below the main sequence, we expect mainly very old, low metallicity sub-dwarfs.
Therefore, we define a magnitude-dependent cut in colour-magnitude space parallel to our chosen segment of the main sequence with slope
\begin{equation}
    \alpha = \frac{M_2 - M_1}{c_2 - c_1}
\end{equation}
where $M_1 = 6 \, \mathrm{mag}$, $M_2 = 7.4 \, \mathrm{mag}$, $c_1 = 0.67 \, \mathrm{mag}$ and $c_2 = 0.9 \, \mathrm{mag}$.
We deem our population sufficiently homogeneous, if all selected stars follow the same number statistics in $z$ independent of their position within the main sequence.
\begin{figure*}
    \centering
    \includegraphics[width=\textwidth]{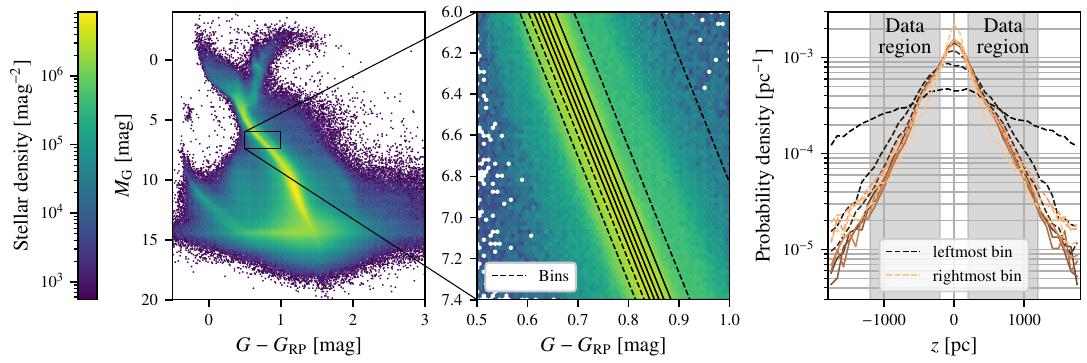}
    \caption{\textit{Left panel:} Colour-magnitude diagram of the initially selected stars in the Gaia DR3 catalogue within our cylindrical volume. \textit{Centre panel:} Zoom-in on the considered region within the main sequence. The lines indicate the edges of our equal-number bins. Every bin, except for the rightmost one, contains $50\,000$ stars. After applying our cuts, only stars in-between the solid lines were kept. \textit{Right panel:} The vertical density profile of the stars in the selected bins. The dashed lines indicate discarded bins. All lines are normalised so that the area under the curve is equal to $1$.}
    \label{fig:ColourMagnitudeDiagram}
\end{figure*}
For that purpose, we perform an equal-numbered binning of stars parallel to the main sequence (bin size $50\,000$ stars), discarding those bins below and above the main sequence that do not share the same statistics modulo sampling noise (see Fig.~\ref{fig:ColourMagnitudeDiagram}).
We note that the cut above the main sequence has only very little effect on the derived results later, while the cut from below is crucial to obtain a homogeneous sample.
This can already clearly be seen in the strongly deviating density profile of the stars in these bins.
The final cuts chosen can be parameterised as
\begin{equation}
    2.02\,\mathrm{mag} \leq M_\mathrm{G} - \alpha (G-G_\mathrm{RP}) \leq 2.27\,\mathrm{mag}\,,
\end{equation}
where $G$ and $G_\mathrm{RP}$ denote the apparent magnitudes in the Gaia broad and red photometric bands, respectively, and $M_\mathrm{G}$ denotes the absolute magnitude in the $G$ band.
After discarding these bins, we are left with $200\,000$ stars.

\subsection{Density profile}
\label{sec:DensityProfile}
From the selected sample, we compute the density profile in the following way:
The number $n_*$ of stars in a given volume as a function of height $z$ is a Poisson process.
This demands a binning of the stars in $z$-space.
This binning is subject to uncertainties because the distances, and therefore the heights, are affected by parallax uncertainties.
Every star in our sample possesses a measurement of the mean and standard deviation of the parallax $p^i$, where $i$ indexes the stars.
We incorporate this uncertainty by sampling from the parallax distribution of all our stars to create $1\,000$ realisations of the density profile.
From these realisations, we compute the mean $N_*^k$ and standard deviation $\sigma_{n_*}^k$ in every bin, where $k$ indexes the $z$-bins.
Since densities are strictly positive, we use a log-normal distribution for describing the uncertainty.
On the one hand, this Gaussian approximation requires a sufficiently large number of stars in each bin.
On the other hand, binning inherently destroys information and larger bins lead to a loss of resolution.
We find a sufficient balance between these two effects using a bin size of $40\,\mathrm{pc}$.
\begin{figure}
    \centering
    \includegraphics[width=\columnwidth]{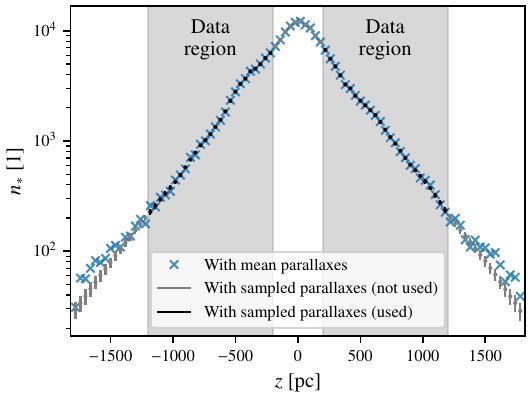}
    \caption{Star counts $n_*$ as a function of height $z$ above the Galactic disk. The blue crosses assume the mean parallax of each star, while the black and grey points show the mean and standard deviation (in log-space) of $1\,000$ realisations sampled from the parallax errors. The bin width is $40\,\mathrm{pc}$.}
    \label{fig:DensityFalloffSampling}
\end{figure} 
The results of this can be seen in Fig.~\ref{fig:DensityFalloffSampling}.
Two effects are important to note: First, using the mean parallax only, the distance is biased towards larger values because parallax and distance are non-linearly related.
Second, the height restriction of the cylinder leads to a loss of stars in the tails of the distribution because by sampling, stars can effectively be moved out of the cylinder while no new stars can enter.
These two effects lead to the divergence of the samples from the raw density profiles at the edges of the cylinder.
We find this effect to be negligible for $|z|\leq1200\,\mathrm{pc}$.
This is because over $98\%$ of these stars have a parallax error smaller than $20\%$, so that the mean parallax is a reasonably good approximation of the true parallax.

\subsection{Stellar velocities}
\label{sec:StellarVelocities}
The Gaia DR3 catalogue \citep{GaiaDR3} contains measurements of the proper motion on the plane of sky for all $200\,000$ stars in our selection, but radial velocities are only available for $154\,217$ stars, mostly at smaller distances and often with large uncertainties.
To extend this towards larger distances, we combine our sample with data from APOGEE \citep{Majewski2017_APOGEE}, GALAH \citep{DeSilva2015_GALAH,Buder2018_GALAH_DR2}, GES \citep{Gilmore2012_GES,Randich2013_GES}, LAMOST \citep{Zhao2012_LAMOST}, and RAVE \citep{Steinmetz2006_RAVE,Steinmetz2020_RAVE1} by means of the Survey-of-Surveys (SoS) catalogue \citep{Tsantaki2022}.
To this end, we cross-match our Gaia DR3 sample to Gaia DR2 source ID's as are provided by the SoS catalogue.
For this, we use the \texttt{dr2\_neighbourhood} table accompanying the Gaia DR3 catalogue.
Our DR3 stars contain a few duplicate crossmatches, where there are multiple DR2 neighbours within the search radius.
We only keep the entry with the smallest angular distance, that is the closest neighbour.
Having cross-matched a one-to-one correspondence, we overwrite Gaia radial velocities with the SoS values where available.
This leads to a total of $160\,888$ stars with radial velocity measurements.
With only Gaia DR3 data, radial velocity data extend only up to approximately $600\,\mathrm{pc}$ above and below the disk.
With the SoS catalogue, this is extended up to all considered heights ($z\leq1800\,\mathrm{pc}$) in positive $z$ direction.
In negative $z$ direction, there is only a minor improvement of statistics and height range.

\section{Methods}
\label{sec:Methods}

\subsection{Poisson-Jeans system of equations}
\label{sec:PoissonJeans}
There are several methods with varied assumptions connecting the statistical properties of a stellar population with the gravitational potential governing their motion \citep{Binney2008}.
We will make use of a particularly simple way, the so-called `minimal assumption method' presented in \citet{Garbari2011} and \citet{Garbari2012} (consult these papers and references therein for details).
We will briefly present the main points in the following.
The collisionless Boltzmann equation describes the time evolution of the phase-space density of some tracer population in a gravitational potential $\Phi$.
Expressing it in terms of the first three moments of the distribution function, we obtain the so-called Jeans equations.
Expressing these in cylindrical coordinates and assuming an axisymmetric system in a steady state, we obtain a system of two equations.
Only the vertical equation is of interest to us, as it connects the total density distribution (by means of the gravitational potential) to the observable velocity moments and density fall-off of the tracer population:
\begin{equation}
    \label{eq:JeansFull}
    \overline{\varv_z^2}\frac{\partial\nu}{\partial z} + \nu\left( \frac{\partial\Phi}{\partial z} + \frac{\partial \overline{\varv_z^2}}{\partial z} + \frac{1}{R\nu}\frac{\partial (R \nu \overline{\varv_z \varv_R})}{\partial R} \right) = 0\,.
\end{equation}
Here, $\nu$ is the density fall-off of some stellar population, $\overline{\varv_z^2}$ and $\overline{\varv_z \varv_R}$ are velocity moments, $z$ is the height above the Galactic disc, $\Phi$ is the gravitational potential, and $R$ is the cylindrical radius.
The last expression in the parentheses is called the `tilt term' and is often neglected (see e.g. \citet{Sivertsson2018_vJeans} or \citet{Guo2020_vJeans} for notable exceptions).
We attempt to improve on this by approximating it as
\begin{equation}
    \label{eq:tiltterm}
    \frac{1}{R \nu}\frac{\partial (\nu R \overline{\varv_z \varv_R})}{\partial R} \approx \frac{\overline{\varv_z \varv_R}}{R} + \frac{\overline{\varv_z \varv_R}}{R_0} + \frac{\overline{\varv_z \varv_R}}{R_\text{T}}
\end{equation}
assuming that $\nu$ and $\overline{\varv_z \varv_R}$ change on some characteristic length scales $R_0$ and $R_\mathrm{T}$.
One would naively expect these scales to be comparable to the disk length scale which is of the order of $2$-$3\,\mathrm{kpc}$ in magnitude.
We compare our expectations to the data by splitting our cylinder in two halves, one closer to the Galactic centre, and one further away, using the mean parallaxes and velocities for estimating these length scales.
We find a sufficient agreement for $R_0$ being approximately constant over the considered range in $z$ and consistent with the disk length scale.
For $R_\mathrm{T}$ however, an estimate is not reliably possible as regions with good velocity statistics (small $|z|$) have negligible $|\overline{\varv_z \varv_R}|$, making the length scale essentially meaningless, and regions with large $|z|$ (and expected large $|\overline{\varv_z \varv_R}|$) have too few stars to yield a useful estimate.
For a more detailed discussion about the sign and magnitude of these corrections, see \citet{Sivertsson2018_vJeans} or \citet{Binney2014}.

\citet{Sivertsson2022} analyse the impact of the tilt term using two simulated Milky-Way-like galaxies out of equilibrium and find that careful modelling of the tilt term is necessary for obtaining an accurate estimate of the local dark matter density.
They further find that, surprisingly, time-dependent effects are subdominant even if the disk experienced a recent perturbation.

With this approximation, the differential equation can be integrated to yield
\begin{equation}
    \label{eq:VerticalDensityFalloff}
    \frac{\nu(z)}{\nu(z_0)} = \frac{\overline{\varv_z^2}(z_0)}{\overline{\varv_z^2}(z)}\exp\left( -\int_{z_0}^{z} \frac{1}{\overline{\varv_z^2}}\frac{\partial\Phi}{\partial z'} + \frac{\overline{\varv_z \varv_R}}{\overline{\varv_z^2}}\left(\frac{1}{R}+\frac{1}{R_0}+\frac{1}{R_\mathrm{T}}\right)\diff z'  \right)\,,
\end{equation}
with some arbitrary reference height $z_0$ which we will take to be $z_0=0$ in the following.

Every tracer population evolving in the gravitational potential $\Phi$ has to satisfy this equation.
The gravitational potential is further constrained by the Poisson equation, here in cylindrical coordinates:
\begin{equation}
    \label{eq:Poisson}
    \frac{\partial^2 \Phi}{\partial z^2} = 4\pi G\left( \rho_\mathrm{s}(z) + \rho_\mathrm{dm}^\mathrm{eff} \right)\,,
\end{equation}
where we have split the total matter density $\rho$ into a visible part $\rho_\mathrm{s}$ and a dark matter part $\rho_\mathrm{dm}^\mathrm{eff}$ with
\begin{equation}
    \label{eq:EffectiveDM}
    \rho_\mathrm{dm}^\mathrm{eff} = \rho_\mathrm{dm} - \frac{1}{4\pi G R}\frac{\partial \varv_\mathrm{c}^2(R)}{\partial R}
\end{equation}
being the effective dark matter density.
It contains the local dark matter density (that we assume to be independent of height $z$ within the considered volume), and another term given by the local slope of the rotation curve ${\varv_\mathrm{c} = (R\partial\Phi/\partial R)^\mathrm{\nicefrac{1}{2}}}$ but independent of $z$.
This term can be calculated from the Oort constants $A$ and $B$ \citep{Binney2008} as
\begin{equation}
    \label{eq:EffDMOortConstants}
    \frac{1}{4\pi G R}\frac{\partial \varv_\mathrm{c}^2(R)}{\partial R} = \frac{1}{2\pi G}\left( B^2 - A^2 \right)\,.
\end{equation}
using the values ${A = 14.79\pm0.11\,\mathrm{km\,s^{-1}\,kpc^{-1}}}$ and ${B = -13.73\pm0.11\,\mathrm{km\,s^{-1}\,kpc^{-1}}}$ as computed by \citet{Akhmetov2024}.

\subsection{Variational Inference}
\label{sec:VariationalInference}
Our main interest lies in one single scalar quantity, the local dark matter density $\rho_\mathrm{dm}$.
It appears in the Poisson equation~(\ref{eq:Poisson}) as one constituent of the mass distribution which dictates the gravitational potential.
This is connected via equation~(\ref{eq:VerticalDensityFalloff}) to the vertical density fall-off $\nu$ of some tracer population and its velocity moments $\overline{\varv_z^2}$ and $\overline{\varv_z \varv_R}$.
The latter two are only indirectly observable and have to be inferred from the data.
To make matters worse, all measurements and input parameters (such as the disc mass model) come with significant uncertainties, and the data is both noisy and (in the case of the stellar velocities) sparse.
This prevents us from using simple histograms to estimate the density fall-off and velocity moments.
The quantities $\overline{\varv_z^2}$ and $\overline{\varv_z \varv_R}$ are generally functions of height $z$ that, as properties of a smooth stellar distribution, can be assumed to be somewhat smooth but are not strongly constrained by data beyond a height of a couple of hundred parsecs.
We will use Gaussian processes to model these functions and capture their uncertainties while exploiting the concept of local correlations.
This allows us to properly incorporate measurement uncertainties, which is a big improvement compared to fitting a function to binned data.
This description inevitably leads to a very large number of latent parameters requiring a sophisticated inference method as sampling, using e.g. MCMC algorithms, is becoming prohibitively costly.

For properly incorporating measurement uncertainties of sparse, noisy data we employ a probabilistic approach.
In order to deal with the large number of latent parameters, we use a novel variational inference method called `Geometric Variational Inference' (geoVI) \citep{geoVI}, a generalisation of `Metric Gaussian Variational Inference' (MGVI) \citep{MGVI}.
It approximates the posterior probability distribution by a multivariate normal distribution in a transformed coordinate system induced by the Fisher information metric.
It has been shown to be a powerful tool for Bayesian inference in high-dimensional parameter spaces and has recently been applied to a variety of problems in astrophysics \citep[e.g.][]{Roth2023,Hutschenreuter2024,Tsouros2024,Edenhofer2024,Soeding2025}.
The algorithm works in an iterative manner by switching between sampling around an expansion point according to the local curvature of the posterior, and updating said expansion point by minimising the Kullback-Leibler-divergence of the approximated posterior and the true posterior until a self-consistent solution is obtained.
The resulting samples trace the true posterior and implicitly contain the correlations between all latent parameters.
The algorithm is implemented in the Python package \texttt{NIFTy8} \citep{Selig2013_nifty1,Steininger2019_nifty3,Arras2019_nifty5,Edenhofer2024_NIFTYRE}.

To perform Bayesian inference, we have to set up a prior, a likelihood, and a forward model.

We incorporate the measurement uncertainties on the parallaxes $p^i$, the proper motions $\varv_\mu^i$ and $\varv_\delta^i$, and the radial velocities $\varv_r^i$ of all stars (indexed by $i$) by sampling from the respective distributions in the prior.
In the forward model, we then compute the vertical velocity $\varv_z^i$, the cylindrical radial velocity $\varv_R^i$, and the vertical position $z^i$ of each star.
These will later provide information on the velocity moments $\overline{\varv_z^2}$ and $\overline{\varv_z \varv_R}$.
Note that even if the errors in parallax-proper-motion space are uncorrelated, the errors in the cylindrical velocities will be correlated.
This is due to projection effects, but also, more importantly, because the parallax to distance transformation is non-linear.

Here, we add a new assumption, namely that the statistics of the velocities of our sample stars in cylindrical coordinates can locally be well-described by a multivariate Gaussian distribution.
Since we do not incorporate angular velocities, this reduces to a 2-dimensional Gaussian distribution in the $(\varv_z, \varv_R)$-plane.
The steady-state and axisymmetry assumptions require that the mean velocities $\overline{\varv_z}$ and $\overline{\varv_R}$ are zero.
The covariance matrix of the velocity distribution is best modelled in its eigenspace.
The two eigenvalues $\Sigma_1$ and $\Sigma_2$ of the covariance matrix $\Sigma$ are equal to $\overline{\varv_z^2}$ and $\overline{\varv_R^2}$ if the matrix is diagonal.
Inspection of the velocity data shows that the velocity moments as a function of height $z$ can be well modelled by a linear function of $|z|$, plus deviations.
We model the deviations using a lognormal Gaussian process by means of the \texttt{NIFTy8} correlated field model \citep{Arras2022}.
This provides the necessary flexibility for the inference to capture the level of smoothness without enforcing a specific functional form or preferred correlation length by hand.
Our model for the eigenvalues then takes the form
\begin{align}
    \label{eq:EVPriors}
    \Sigma_1 (z) & = \left(b_1 + m_1 |z|\right) \exp\left( g_{\Sigma_1}(z) \right)     \\
    \Sigma_2 (z) & = \left(b_2 + m_2 |z|\right) \exp\left( g_{\Sigma_2}(z) \right) \,,
\end{align}
where $g_{\Sigma_1}$ and $g_{\Sigma_2}$ are the Gaussian processes, $b_1$ and $b_2$ are the intercepts, and $m_1$ and $m_2$ are the slopes of the linear relations.

We expect - and will allow for - deviations from a diagonal covariance matrix by introducing an angle $\theta$ rotating the covariance matrix.
Our prior consists of two parts: First, a tilt originating from rotations relative to the Galactic centre.
This is expected to be the majority contribution, though may not fit exactly.
To give the angle more freedom, we add a second Gaussian process $g_\theta$ that is added to a uniformly distributed (arbitrary) starting angle $\theta_0$:
\begin{equation}
    \label{eq:AnglePrior}
    \theta(z) = -\arctan\left(\frac{z}{R_\odot}\right) + \theta_0 + \left(g_\theta(z) - g_\theta(0)\right) \,.
\end{equation}
The covariance matrix is then assembled as
\begin{equation}
    \Sigma(z) = R(\theta(z)) \begin{pmatrix}
        \Sigma_1(z) & 0           \\
        0           & \Sigma_2(z)
    \end{pmatrix} R(\theta(z))^\mathrm{T} \,.
\end{equation}
with a 2D rotation matrix $R(\theta)$.
The components describe the velocity moments, that is $\Sigma_{00} = \overline{\varv_z^2}$, $\Sigma_{01} = \Sigma_{10} = \overline{\varv_z \varv_R}$, and $\Sigma_{11} = \overline{\varv_R^2}$.
We demand the stellar velocities to be distributed according to this covariance matrix, that means that the residuals of the velocities are distributed according to a multivariate Gaussian distribution with zero mean and covariance matrix $\Sigma$.
This constitutes our first likelihood:
\begin{equation}
    \renewcommand\arraystretch{1.3}
    \mathrm{LH}_1=\mathscr{G}\left(\begin{pmatrix}\varv_z^i\\ \varv_R^i\end{pmatrix}\;\middle\vert\; \vec{0}, \Sigma(z)\right)\,.
\end{equation}

Now that we have a model for the velocity moments, we can turn to the vertical density fall-off $\nu(z)$.
For computing it, using equation~(\ref{eq:VerticalDensityFalloff}), we need to know the first derivative of the gravitational potential $\Phi$.
We can compute this by integrating the Poisson equation~(\ref{eq:Poisson}) with respect to $z$ using the boundary condition
\begin{equation}
    \Phi(0) = \frac{\partial\Phi}{\partial z}(0) = 0\,.
\end{equation}
The right-hand side of the Poisson equation is given by the sum of the visible mass density $\rho_\mathrm{s}$ (see equation~(\ref{eq:IsothermalMassModel})) and the effective dark matter density $\rho_\mathrm{dm}^\mathrm{eff}$.
We sample the constituents of the former from the values given in Table~\ref{tab:massmodel}, the dark matter density from a uniform prior between $0\,\mathrm{M}_\odot \mathrm{pc}^{-3}$ and $0.1\,\mathrm{M}_\odot \mathrm{pc}^{-3}$, and the Oort's constants from a normal prior according to their uncertainties (cf. equations (\ref{eq:EffectiveDM}) and (\ref{eq:EffDMOortConstants})).

Next, we need the scale lengths $R_0$ and $R_\mathrm{T}$ for the tilt term (equation~(\ref{eq:tiltterm})).
For the number density scale length $R_0$, we choose a Gaussian prior centred around $-2.5\,\mathrm{kpc}$ with a standard deviation of $0.5\,\mathrm{kpc}$.
For the velocity tilt scale length $R_\mathrm{T}$, there is no conclusive data available that confirms the magnitude, constancy, or even the sign of this term.
Indications of disequilibria in the velocity distribution that would lead to a contribution that is potentially large in magnitude and varying in height $|z|$ have been known for a long time \citep[e.g.][]{Dehnen1998}.
We therefore model this term as a Gaussian process as well, choosing a broad prior centred around zero.
Another option would be to choose a scalar Gaussian prior, assuming that the value is constant in $z$.
To estimate the effect of such a prior choice, we will also discuss results derived using a scalar prior.

With these ingredients, equation~(\ref{eq:VerticalDensityFalloff}) can be evaluated to yield the vertical density fall-off $\nu(z)$.
To account for the uncertainties in our data due to parallax errors, we multiply this density fall-off by the above-derived (see section \ref{sec:DensityProfile}) star count error $\sigma_{n_*}^k$ in log-space to obtain the density profile $n_*(z)$ in every bin $k$ as
\begin{equation}
    n_*^k = n_{*,0} \nu(z^k) \exp\left(\sigma_{n_*}^k\right) \,.
\end{equation}
Here, $n_{*,0}$ is the density at the reference height ${z_0=0\,\mathrm{pc}}$, which is a-priori unknown.
We draw it from a broad log-normal prior around the actually measured density $N_{*,0}$ (which we cannot take directly because data in this region might be volume-incomplete).
This can then be compared to the measured star counts $N_*^k$ using a Poissonian likelihood:
\begin{equation}
    \mathrm{LH}_2 = \mathscr{P}\left(N_*^k \;\middle\vert\; n_*^k\right)\,.
\end{equation}
This procedure captures both the stochastic Poissonian uncertainties in the star counts and the parallax-induced uncertainties in the density fall-off.

Finally, we include the constraints on the surface mass density within $1.1\,\mathrm{kpc}$ of stellar and gaseous components of the disk mass model from \citet{McKee2015_vJeans} as two additional scalar likelihoods (cf. section~\ref{sec:DiscMassModel}):
\begin{align}
    \label{eq:LH3LH4}
    \mathrm{LH}_3 & = \mathscr{G}\left(\Sigma_{\mathrm{g},1.1}\;\middle\vert\; 12.6\pm1.6\,\mathrm{M}_\odot\,\mathrm{pc}^{-2}\right) \\
    \mathrm{LH}_4 & = \mathscr{G}\left(\Sigma_{\mathrm{*,1.1}}\;\middle\vert\; 31.2\pm3.0\,\mathrm{M}_\odot\,\mathrm{pc}^{-2}\right) \,.
\end{align}
These constraints help in alleviating some degeneracies between baryonic and dark matter components in the disk mass model.

\section{Results}
\label{sec:Results}
We ran the variational inference algorithm using 100 antithetical (200 total) samples for 12 iterations.
We report a mean reduced $\chi^2$ (per degree of freedom, averaged over all samples) of $1.0$ for the fit indicating that all data points included are statistically consistent with the model.
This includes the Poissonian star counts and the multivariate Gaussian velocity data.

The main difference setting our work apart from others, is our use of Gaussian processes to model a-priori unknown but smooth functions and the variational inference algorithm to sample the posterior distribution.
We will first discuss the inferred velocity moments and the fit to the density fall-off, and then the estimated local dark matter density and its dependence on modelling of the tilt term.

There are three velocity moments that we can compute from the inferred covariance matrix: the vertical velocity dispersion $\overline{\varv_z^2}$, the radial velocity dispersion $\overline{\varv_R^2}$, and the cross term $\overline{\varv_z \varv_R}$.
\begin{figure}
    \centering
    \includegraphics[width=\columnwidth]{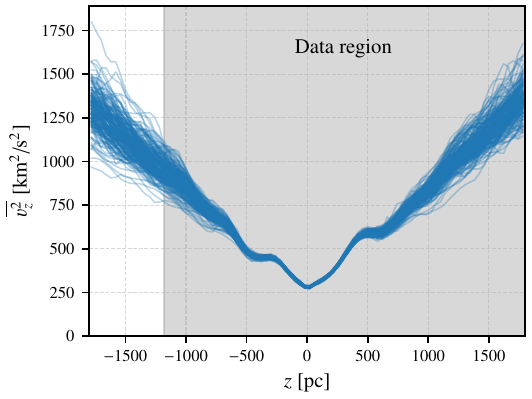}
    \caption{Vertical velocity dispersion $\overline{\varv_z^2}$ as a function of height $z$ above the Galactic disk. All samples are shown. The shaded region indicates the availability of stellar velocity data.}
    \label{fig:resvz2}
\end{figure}
The inferred vertical velocity dispersion is presented in Fig.~\ref{fig:resvz2}.
It shows the approximately linear relation with $|z|$ that we expected.
Within the data region, especially towards small $|z|$, the function is tightly constrained by the data.
Towards larger $|z|$, where data becomes sparse, the uncertainties increase significantly as indicated by the spread of samples.
We observe an asymmetry in the vertical velocity dispersion above and below the disk (see e.g. the bump at around $z=\pm400\,\mathrm{pc}$), even in regions well-constrained by data.
This hints towards one of our assumptions being violated, e.g. the assumption of axisymmetry or steady-state.
In the density fall-off data (see Fig.~\ref{fig:DensityFalloffSampling}), a similar asymmetry is observed at the same heights, but opposite in sign.
This is somewhat expected from the solution of the Jeans equation~(equation~(\ref{eq:VerticalDensityFalloff})), which relates the density fall-off $\nu$ and the vertical velocity dispersion $\overline{\varv_z^2}$ approximately antilinearly.
As the gravitational potential $\Phi$ is enforced to be smooth and symmetric, the asymmetries in the other terms have to, and do, cancel each other (within the uncertainty margins) to match the observed star-count and velocity data.

\begin{figure}
    \centering
    \includegraphics[width=\columnwidth]{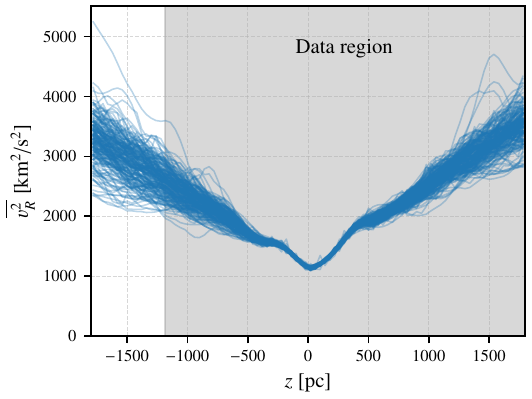}
    \caption{Like Fig.~\ref{fig:resvz2}, but for the radial velocity dispersion $\overline{\varv_R^2}$.}
    \label{fig:resvr2}
\end{figure}
The radial velocity dispersion $\overline{\varv_R^2}$ is shown in Fig.~\ref{fig:resvr2}.
It is generally larger than the vertical velocity dispersion and also shows an approximately linear relation with $|z|$.
A similar asymmetrical behaviour as for the vertical velocity dispersion is observed, but this function is generally less constrained by the data.
This may be due to the fact, that only the stellar velocity data directly contributes here, while in case of the vertical velocity dispersion, also the star counts contribute indirectly (by means of equation~(\ref{eq:VerticalDensityFalloff})).
Especially at negative $z$, where there is no velocity data, many samples deviate significantly from the mean, reflecting the lack of velocity data in this region.

\begin{figure}
    \centering
    \includegraphics[width=\columnwidth]{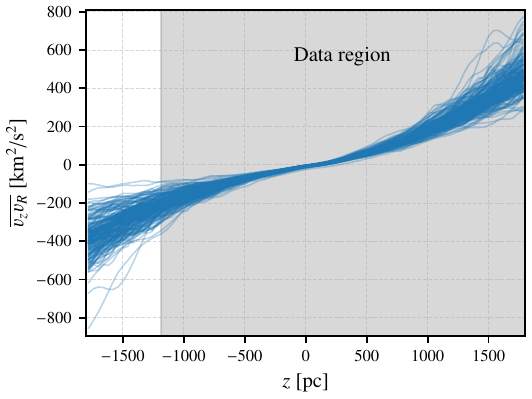}
    \caption{Like Fig.~\ref{fig:resvz2}, but for the velocity covariance $\overline{\varv_z \varv_R}$.}
    \label{fig:resvzvr}
\end{figure}
The cross term $\overline{\varv_z \varv_R}$ is shown in Fig.~\ref{fig:resvzvr}.
It follows a comparatively smoother behaviour than the other two velocity moments and is on average about an order of magnitude smaller.
It exhibits, again, a wide spread at large distances and is close to zero in the mid-plane.
The gradient at $z=0\,\mathrm{pc}$ is positive which indicates that on both sides of the disk, stars that are moving vertically outwards, are also preferentially moving radially outwards.
A simple tilt in the average stellar orbits leads to the expectation of $\overline{\varv_z \varv_R}$ being antisymmetric in $z$, which we confirm within the uncertainty margins.
At least in regions well-constrained by data, so at small $|z|$, the cross term, and also the tilt angle, increases, indicating that the correlation of velocity components becomes increasingly important for analyses of stars towards larger distances from the disk.
However, velocity data is sparse in these regions, leading to large uncertainties.
Within our error estimate, the inferred tilt is consistent with a simple rotation of the stellar orbits around the Galactic centre.

\begin{figure}
    \centering
    \includegraphics[width=\columnwidth]{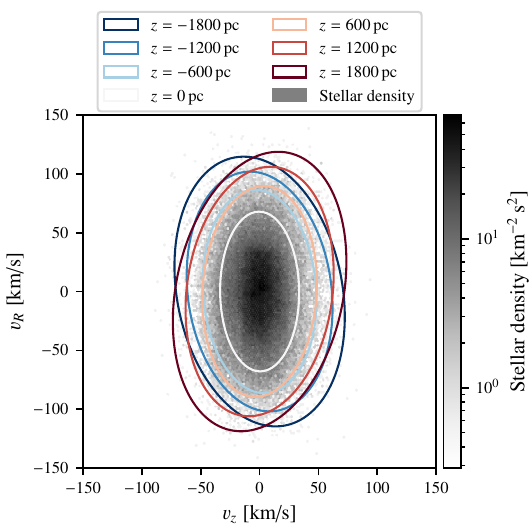}
    \caption{Density map of stellar velocities in the $(\varv_z, \varv_R)$-plane in our dataset. The sample-mean of the inferred covariance matrix is indicated by the 2-$\sigma$ ellipse contours for various heights $z$.}
    \label{fig:covtilt}
\end{figure}
The covariance matrix of the stellar velocities is shown in Fig.~\ref{fig:covtilt} for various $z$.
This emphasises the tilt of the covariance matrix in the considered 2-dimensional velocity space.
It should be noted that the spread of the samples, that reflects the uncertainty of the covariance matrix estimation, is very large for the $z=\pm1800\,\mathrm{pc}$ samples.
At these distances, it is only constrained by extrapolation from smaller $|z|$ and our prior structure for the Gaussian processes and angles (cf. equations (\ref{eq:EVPriors}) and (\ref{eq:AnglePrior})).
However, the rotation away from a diagonal covariance matrix is clearly visible, statistically significant, and consistent with findings of previous analyses such as \citet{Buedenbender2015}.

\begin{figure}
    \centering
    \includegraphics[width=\columnwidth]{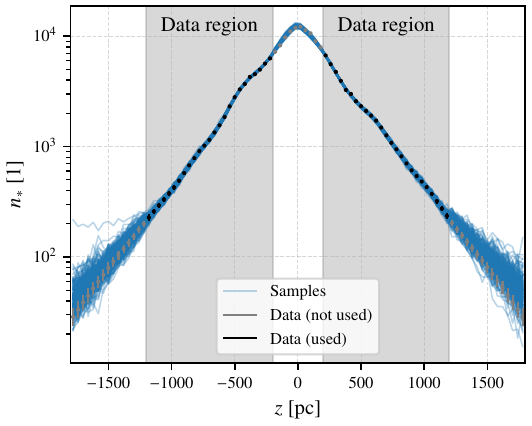}
    \caption{Star counts $n_*$ as a function of height $z$ above the Galactic disk. All samples are shown. The shaded region indicates the availability of star count data.}
    \label{fig:resDensityFalloff}
\end{figure}
The inferred density fall-off $\nu(z)$ is shown in Fig.~\ref{fig:resDensityFalloff}.
The model is in agreement with the data, indicated by a good fit of the star counts in the considered region.
At distances farther than $|z|=1200\,\mathrm{pc}$, so outside the range of star count data considered, our samples show a large spread, reflecting the lack of constraining data.
See the discussion in section \ref{sec:DensityProfile} for an explanation of why data outside this region was excluded.
Similar to the velocity moments, we can also see a clear asymmetry in the density fall-off above and below the disk at approximately $z=\pm400\,\mathrm{pc}$.
Interestingly, these asymmetries fit together very well, being able to explain the asymmetric data even in a symmetric gravitational potential.
It should be emphasised, that these are asymmetries in the tracer population only, whereas the disk mass model is still fully symmetric about the mid-plane.

We find a local dark matter density of
\begin{equation}
    \rho_\mathrm{dm} = 0.0117 \pm 0.0035\, \mathrm{M}_\odot\,\mathrm{pc}^{-3} = 0.44 \pm 0.13\, \mathrm{GeV}\,\mathrm{cm}^{-3}\,,
\end{equation}
which already contains the correction term using Oort's constants.
This result uses the Gaussian process prior for $R_\mathrm{T}^{-1}$ in the tilt term (equation~(\ref{eq:tiltterm})).
This result is consistent with values typically found in the literature \citep[compare with e.g.][]{deSalas2020} for local analyses, but significantly lower than the (mean) value derived by \citet{Garbari2012} using a similar class of stars.
As a consistency check, we also ran our analysis using only data above or below the disk, finding dark matter densities of ${\rho_\mathrm{dm}^\mathrm{north} = 0.0149 \pm 0.0045\, \mathrm{M}_\odot\,\mathrm{pc}^{-3}}$ and ${\rho_\mathrm{dm}^\mathrm{south} = 0.0149 \pm 0.0037\, \mathrm{M}_\odot\,\mathrm{pc}^{-3}}$, which agree with both each other and with the value derived using the full data set.
The total baryonic surface mass density is prior-dominated, with a value of ${\Sigma_\mathrm{b} = 43.75 \pm 3.04 \mathrm{M}_\odot\,\mathrm{pc}^{-2}}$.

\begin{figure}
    \centering
    \includegraphics[width=\columnwidth]{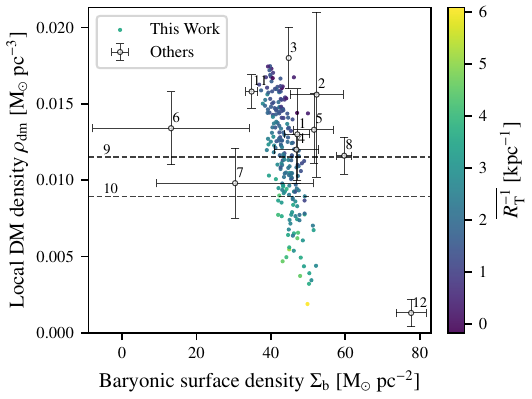}
    \caption{
        Our estimation of the local dark matter density in comparison to recent estimates using a vertical Jeans analysis.
        The colour coding indicates the mean value of $R_\mathrm{T}^{-1}$ of each posterior sample, averaged over the $z$-axis.
        The other works, shown in grey, are \citet{McKee2015_vJeans} (1), \citet{Xia2016_vJeans} (2), \citet{Hagen2018_vJeans} (3), \citet{Sivertsson2018_vJeans} (4), \citet{Guo2020_vJeans} (5), \citet{Salomon2020_vJeans} (6 north and 7 south), \citet{Wardana2020_vJeans} (8), \citet{Nitschai2020_aniJeans,Nitschai2021_aniJeans} (9 and 10), and \citet{Cheng2024} (11 thin disk and 12 thick disk).
        We have omitted one recent result by \citet{LopezCarredoira2025}, because the error bars are so large that readability would suffer.
        For some works, we had to assume a locally constant dark matter density for calculating their baryonic surface mass density.
    }
    \label{fig:resDMOverview}
\end{figure}
In Fig.~\ref{fig:resDMOverview}, we compare our result to other recent Jeans analyses of the local dark matter density.
Apart from consistency with the literature, it also illustrates that the estimated dark matter density is anti-correlated with both the baryonic surface mass density $\Sigma_\mathrm{b}$ and the inverse characteristic length scale $R_\mathrm{T}^{-1}$ of the tilt term.
Deviations between different Jeans analyses can stem from different choices of the tracer population, the particular data used, and the approximations made in the Jeans equation.
In the following, we want to examine the influence of our modelling of the tilt term on the inferred dark matter density.

In order to quantify the effect of our particular modelling of the tilt term, we also ran an analysis with a scalar prior for $R_\mathrm{T}^{-1}$, assuming a constant behaviour in $z$.
\begin{figure*}
    \centering
    \includegraphics[width=\textwidth]{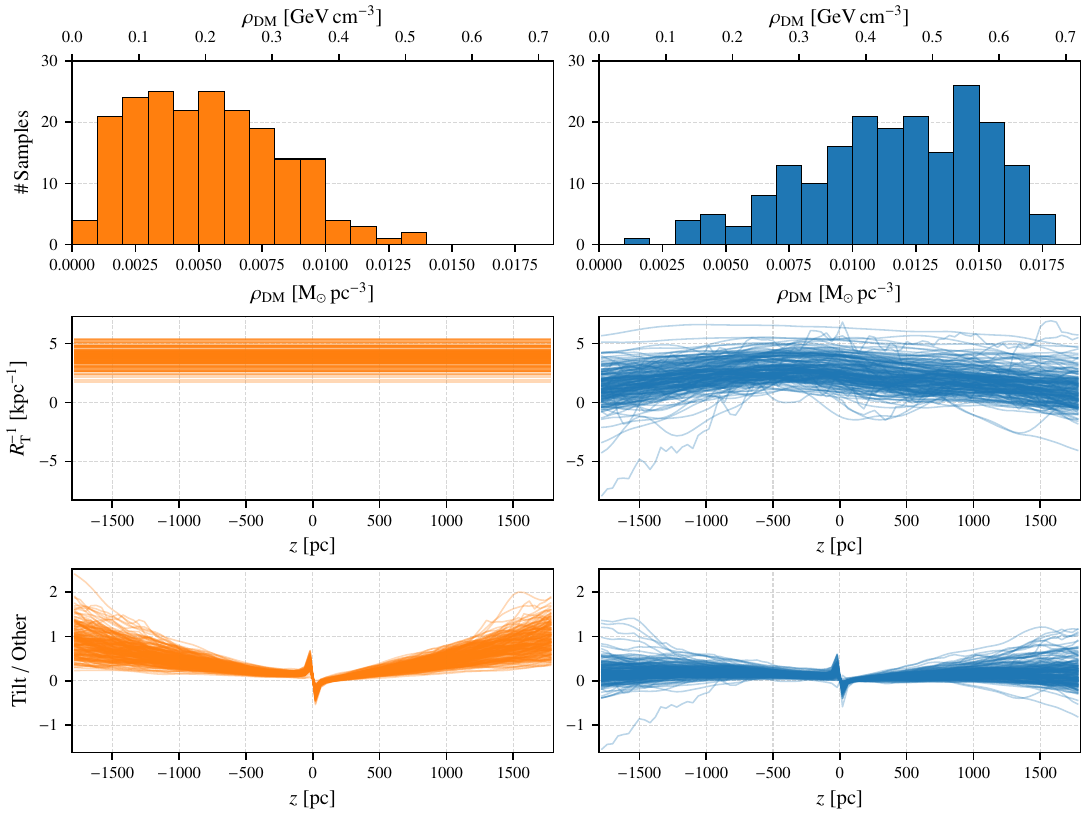}
    \caption{Comparison of the effect of modelling the tilt term on the inferred dark matter density. The left column shows the analysis using a scalar prior for $R_\mathrm{T}^{-1}$, the right column the analysis using a Gaussian process prior. All samples are shown. \textit{Top row:} Histogram showing samples of the inferred posterior distribution of the local dark matter density $\rho_\mathrm{dm}$. \textit{Middle row:} The characteristic length scale $R_\mathrm{T}^{-1}$ as a function of height $z$ above the Galactic disk. \textit{Bottom row:} The relative contribution of the tilt term in the Jeans equation as a function of height $z$ above the Galactic disk. For details, refer to the text.}
    \label{fig:rescompDM}
\end{figure*}
We show a comparison of the inferred dark matter densities as a histogram, $R_\mathrm{T}^{-1}$ as a function of $z$, and the relative contribution of the tilt term in Fig.~\ref{fig:rescompDM}.

Using the scalar prior, we find a local dark matter density of
\begin{equation}
    \rho_\mathrm{dm} = 0.0053 \pm 0.0028\, \mathrm{M}_\odot\,\mathrm{pc}^{-3} = 0.20 \pm 0.11\, \mathrm{GeV}\,\mathrm{cm}^{-3}\,,
\end{equation}
which is significantly lower than the value found using the Gaussian process prior.
Here, we find a difference to the values derived using only data above or below the disk, in particular ${\rho_\mathrm{dm}^\mathrm{north} = 0.0135 \pm 0.0045\, \mathrm{M}_\odot\,\mathrm{pc}^{-3}}$ and ${\rho_\mathrm{dm}^\mathrm{south} = 0.0155 \pm 0.0052\, \mathrm{M}_\odot\,\mathrm{pc}^{-3}}$.
Note that even though the difference may seem large, the uncertainties are also large, and the values are still consistent with each other.
We attribute this difference to the asymmetry in the data. Fitting both sides of the disk simultaneously demands more flexibility in the model, favouring larger values of $R_\mathrm{T}^{-1}$ and thus lower values of $\rho_\mathrm{dm}$.
The baryonic surface mass density is slightly larger than in the Gaussian process prior case, with a value of ${\Sigma_\mathrm{b} = 46.07 \pm 2.70 \mathrm{M}_\odot\,\mathrm{pc}^{-2}}$.

The characteristic length scale $R_\mathrm{T}^{-1}$ of the tilt term is shown in the middle row of Fig.~\ref{fig:rescompDM}.
Somewhat surprisingly, it prefers positive values indicating that the correlated velocity moment $\overline{\varv_z \varv_R}$ is increasing towards larger $R$.
This could be caused either by an increasing tilt angle, or by larger velocity dispersions towards larger $R$.
The scalar prior prefers values of about $3.82 \pm 0.72 \,\mathrm{kpc}^{-1}$, while the Gaussian process prior prefers values of about $2.11 \pm 0.98 \,\mathrm{kpc}^{-1}$ after averaging over $z$.
The preferred values of order unity lead to a net positive contribution of the tilt term to the exponential fall-off in equation~(\ref{eq:VerticalDensityFalloff}), meaning a steepening of the vertical density fall-off $\nu$.
Such a general steepening can also be achieved by increasing the local dark matter density $\rho_\mathrm{dm}$ as this translates to the gravitational force $\frac{\partial\Phi}{\partial z}$ gaining an additional, linear term.
This explains why the dark matter density is anti-correlated with the product of $\overline{\varv_z \varv_R}$ and $R_\mathrm{T}^{-1}$ (cf. equation~(\ref{eq:VerticalDensityFalloff})).
Since the allowed range of $\overline{\varv_z \varv_R}$ is dictated by data which is equal for both priors, the difference in the inferred dark matter density is mainly due to the different values of $R_\mathrm{T}^{-1}$.
We show the relative contribution of the tilt term in the Jeans equation in the bottom row of Fig.~\ref{fig:rescompDM}.
The relative contribution is defined as the ratio of the tilt term to the other term in the Jeans equation~(equation~(\ref{eq:VerticalDensityFalloff})), in particular
\begin{equation}
    \label{eq:resTiltTerm}
    \frac{\text{Tilt}}{\text{Other}} = \frac{\overline{\varv_z \varv_R}\left(\frac{1}{R}+\frac{1}{R_0}+\frac{1}{R_\mathrm{T}}\right)}{\frac{\partial\Phi}{\partial z}}\,.
\end{equation}
As both the numerator and denominator change sign near $z=0\,\mathrm{pc}$, the sharp peaks and change of sign there are not of interest.

For the Gaussian process prior, the tilt term contributes less than $30\%$ of the other term in the Jeans equation over the whole $z$ region considered for most samples.
Especially towards small $|z|$, the tilt term is small, as $\overline{\varv_z \varv_R}$ is close to zero there.
There is a large spread around the mean, where, for some samples, the tilt term is the dominant contribution at large $|z|$.

For the scalar prior, the tilt term contributes much more on average.
It has a net positive contribution for all samples and at all $z$, except for the very small $|z|$ region.
Since these large values of $R_\mathrm{T}^{-1}$ are only obtained in the region around $z=-400\,\mathrm{pc}$ for the Gaussian process prior, this region might be responsible for dictating the large value everywhere in case of the scalar prior.
This enforces a significant contribution of the tilt term everywhere, increasing its importance as a consequence.
The assumption of separability of the radial and vertical direction in the tilt term (equation~(\ref{eq:tiltterm})) is not well justified and likely too restrictive.
This makes the tilt term important for the Jeans equation at all $z$ considered, contributing more than $50\%$ of the other term in the Jeans equation for most samples at large $|z|$.

Generally, the tilt term is not well-constrained, consequence of sparse velocity data at large $z$, and the lack of explicit data and modelling.
This result is consistent with the findings of \citet{Sivertsson2018_vJeans} who find a strong tension of prior and posterior for this kind of approximation in one of their stellar sample populations (using G-dwarfs).
They make use of a uniform prior for $\frac{1}{R_0}+\frac{1}{R_\mathrm{T}}$ restricted to values smaller than those found here.
Assuming for a moment that the value we found is physically correct, it is not surprising that they found a strong tension.
Whether there really is a radial increase of $\overline{\varv_z \varv_R}$, as implied by the positive value of $R_\mathrm{T}^{-1}$, is something only the data can tell.
Unfortunately, the current data lack the statistical power to allow a definitive conclusion.
It may well be that this approach to approximating the Jeans equations and parametrising the tilt term is problematic, as it can compensate for broken assumptions (e.g. steady state or axisymmetry) by absorbing their effects into the tilt term.
We however find an excellent data fit with our model, enabled mainly by the flexibility of the Gaussian processes.

We conclude that the tilt term is, at least in our approximation using a Gaussian process as prior, subdominant but not negligible, contributing less than approximately $50\%$ relative to the other term in the Jeans equation at all heights $z$ considered in this analysis and for most samples.

\section{Conclusions}
\label{sec:Conclusions}
In this work, we present a new analysis of the local dark matter density in the Galactic disk using a sample of stars from the Gaia DR3 \citep[][]{GaiaDR3} catalogue with available radial velocity measurements extended using the Survey-of-Surveys \citep[SoS,][]{Tsantaki2022} catalogue.
The analysis was performed using geometric variational inference \citep[geoVI,][]{geoVI} implemented in the \texttt{NIFTy8} package.
The velocity moments are modelled as Gaussian processes in the eigenspace of the 2D covariance matrix of radial and vertical velocities.
We also model the tilt term in the Jeans equation as a Gaussian process in $z$, avoiding the need for assuming separability of the radial and vertical components, or a restrictive functional form.
We find a local dark matter density of ${\rho_\mathrm{dm} = 0.0117 \pm 0.0035\, \mathrm{M}_\odot\,\mathrm{pc}^{-3}}$ which is in agreement with previous, similar estimates.
Comparing this to an analysis using a scalar prior for the tilt term, we demonstrate that the modelling of the tilt term has a significant effect on the inferred local dark matter density and can lead to severe biases.

This work serves not only as an updated estimation of the local dark matter density (which has been performed many times before), but also as a proof of concept for two things: First, the application of Gaussian processes to model the velocity moments of a Galactic tracer population as a function of $z$, and second, the ability to infer nondiagonal elements of a multivariate Gaussian likelihood within this family of variational inference methods.
The first point naturally leads to a steep increase in the number of latent parameters, requiring moving away from MCMC-based sampling methods of sampling the posterior.
The application of variational inference methods also allows us to include measurement uncertainties in a much more natural way than usually done, that is not by Gaussian error propagation, but by sampling the uncertainties directly in the prior and then transforming them to the desired physical quantities within the forward model.
This, too, increases the number of effective latent parameters by a large factor, further manifesting the need for an efficient inference method.

We confirm the smooth, approximately linear, but not featureless nature of the velocity moments as a function of height $z$.
Furthermore, we find a significant correlation of radial and vertical stellar velocities (manifesting in a non-zero off-diagonal term in the covariance matrix), increasing with $|z|$ as well, which is consistent with previous findings \citep[compare with e.g.][]{Buedenbender2015}.
Finally, the effect of the tilt term is quantified using a local approximation and found to be subdominant but not negligible in the Jeans equation, contributing less than $50\%$ of the other term in the Jeans equation for all heights considered in this analysis and for most samples.
Please note that most samples have a much smaller contribution, especially at small $|z|$.

For improving upon this in future analyses, there are multiple options: First, the mass model of the baryonic components can be improved by allowing for more freedom, especially an asymmetry in the mass distribution above and below the disk.
The fact that there is an apparent asymmetry in the tracer population (here, and also in previous analyses using different families of tracers) implies the possibility of an asymmetry in the overall mass distribution, especially for large $|z|$.
Second, the spatial dimensionality of the problem can be increased in various ways.
The assumption of axisymmetry can be removed by adding the angular velocity as a third component to the velocity covariance.
This change would likely go hand in hand with an increase of spatial dimensionality, i.e. making the velocity moments and stellar density not only a function of height $z$, but also of the galactocentric radius $R$ and azimuthal angle $\varphi$.
This would give direct access to the required spatial derivatives.
As a third, and possibly the most difficult, avenue for improvement, the time axis can be considered.
Essentially all past analyses, apart from simulations, assume a steady-state system.
This assumption is being increasingly questioned as evidence for disequilibria, excited for example by spiral arms, stellar streams, or the Large Magellanic Cloud, is mounting \citep[e.g.][among many others]{Widrow2012,Hou2014,Antoja2018,Laporte2018,Vasiliev2021}.
All of these extensions to the model would lead to a significant increase in model complexity and computational requirements.
Modelling using Gaussian processes also has some downsides, especially when simultaneously neglecting relevant parts of the model (e.g. the time dependence), since the freedom of the Gaussian processes can absorb effects that belong in a different, unmodelled field.
Thus, while we strongly recommend their use, we also recommend caution when interpreting the results.

We conclude by stressing that the methods used here are very versatile and can be applied to a wide range of problems, even beyond Galactic astrophysics.
The combination of sophisticated inference methods and the ever-increasing amount of large, high quality datasets may soon lead to a new era of data-driven, high-dimensional modelling in astrophysics.

\section*{Acknowledgements}
We would like to thank João Alves for helpful comments regarding Gaia data.
All figures in this publication have been created
using matplotlib \citep{Hunter2007_matplotlib}.
Part of this work was supported by the German Deutsche Forschungsgemeinschaft, DFG project number 495252601.
The authors gratefully acknowledge the computing time provided to them at the NHR Center NHR4CES at RWTH Aachen University (project number p0021785). This is funded by the Federal Ministry of Education and Research, and the state governments participating on the basis of the resolutions of the GWK for national high performance computing at universities (\url{https://www.nhr-verein.de/unsere-partner}).
Part of this work is based on archival data, software or online services provided by the Space Science Data Center - ASI. In particular, the GaiaPortal access tool was used for this research (\url{http://gaiaportal.ssdc.asi.it}).

\section*{Data Availability}
The Gaia DR3 data has been downloaded using the python package \texttt{astroquery} \citep{Ginsburg2019_astroquery}.
The Survey-of-Surveys (SoS) data was downloaded from the GaiaPortal website (\url{https://gaiaportal.ssdc.asi.it/SoS/query/form}).
The processed datasets and posterior samples will be shared on reasonable request to the corresponding author.

\bibliographystyle{mnras}
\bibliography{references}


\bsp	
\label{lastpage}
\end{document}